\documentclass[conference]{IEEEtran}
\usepackage{cite}
\usepackage{graphicx}
\usepackage{booktabs}
\usepackage{array}
\usepackage{multirow}
\usepackage{makecell}
\usepackage{pgfplots}
\pgfplotsset{compat=1.17}

\ifCLASSINFOpdf
\else
\fi
\usepackage{multirow}

\begin{document}
%
\title{Sentiment Analysis in Software Engineering: Evaluating Generative Pre-trained Transformers}

\author{\IEEEauthorblockN{KM Khalid Saifullah}
\IEEEauthorblockA{Email: tsaifullah25@wooster.edu}
\and
\IEEEauthorblockN{Faiaz Azmain}
\IEEEauthorblockA{Email: fazmain25@wooster.edu}
\and
\IEEEauthorblockN{Habiba Hye}
\IEEEauthorblockA{Email: haabdulhye25@wooster.edu}}


%


\maketitle



\begin{abstract}
Sentiment analysis plays a crucial role in understanding developer interactions, issue resolutions, and project dynamics within software engineering (SE). While traditional SE-specific sentiment analysis tools have made significant strides, they often fail to account for the nuanced and context-dependent language inherent to the domain. This study systematically evaluates the performance of bidirectional transformers, such as BERT, against generative pre-trained transformers, specifically GPT-4o-mini, in SE sentiment analysis. Using datasets from GitHub, Stack Overflow, and Jira, we benchmark the models' capabilities with fine-tuned and default configurations. The results reveal that fine-tuned GPT-4o-mini performs comparable to BERT and other bidirectional models on structured and balanced datasets like GitHub and Jira, achieving macro-averaged F1-scores of 0.93 and 0.98, respectively. However, on linguistically complex datasets with imbalanced sentiment distributions, such as Stack Overflow, the default GPT-4o-mini model exhibits superior generalization, achieving an accuracy of 85.3\% compared to the fine-tuned model's 13.1\%. These findings highlight the trade-offs between fine-tuning and leveraging pre-trained models for SE tasks. The study underscores the importance of aligning model architectures with dataset characteristics to optimize performance and proposes directions for future research in refining sentiment analysis tools tailored to the SE domain.
\end{abstract}
\begin{IEEEkeywords}
Sentiment analysis, Software engineering, Transformer models, Generative pre-trained transformers, BERT, Fine-tuning, Natural language processing, Gold-standard datasets.
\end{IEEEkeywords}


%
\IEEEpeerreviewmaketitle

\section{Introduction}

Sentiment analysis, a critical subfield of natural language processing (NLP), involves classifying text into sentiment polarities, such as positive, neutral, and negative. It has been widely studied across various domains, including software engineering (SE), where analyzing sentiments expressed in textual artifacts provides insights into developer interactions, issue resolution, and project dynamics. For instance, sentiments in GitHub pull requests, commit comments, and Stack Overflow discussions have been shown to influence collaboration and productivity in SE projects \cite{medhat14}\cite{islam18}. Despite its significance, sentiment analysis in SE faces unique challenges, including technical jargon, domain-specific expressions, and informal communication patterns, which make general-purpose sentiment analysis tools inadequate \cite{choudhury13}.

The development of SE-specific sentiment analysis tools has attempted to address these challenges. Early tools like SentiStrength \cite{thelwall10} and its SE-specific variant SentiStrength-SE \cite{ahmed17} tailored lexicon-based approaches to the SE domain by incorporating domain-specific dictionaries and rules. Supervised learning tools such as Senti4SD and SentiCR further enhanced sentiment classification by leveraging n-gram features and machine learning techniques \cite{ahmed17}\cite{novielli2018benchmark}. However, these tools often struggled to handle nuanced or context-dependent language, highlighting the need for more sophisticated approaches.

Recent advancements in NLP have introduced pre-trained transformer-based models, such as Bidirectional Encoder Representations from Transformers (BERT) \cite{devlin18}, RoBERTa \cite{liu19}, and XLNet \cite{yang19}, which leverage contextual word representations to significantly improve performance across various tasks. These models, pre-trained on large corpora, can be fine-tuned for downstream tasks like sentiment analysis, making them particularly attractive for SE applications \cite{qiu20}. Unlike earlier approaches, transformer models dynamically capture the context of words within sentences, addressing the limitations of static embeddings and lexicon-based methods \cite{qiu20}\cite{devlin18}.

Building on these advancements, Zhang et al. demonstrated that fine-tuned transformer models significantly outperformed traditional SE-specific tools like SentiStrength-SE and Senti4SD, achieving improvements of up to 35.6\% in macro-averaged F1 scores. \cite{zhang2020} While Zhang et al.'s work laid the groundwork for transformer-based SA4SE, their study focused exclusively on bidirectional transformers, such as BERT and its variants, without considering generative transformer models like GPT (Generative Pre-trained Transformer). Generative transformers, characterized by their autoregressive architectures, have shown remarkable performance in generating coherent sequences and processing large-scale text data \cite{qiu20}. However, the comparative efficacy of bidirectional transformers and generative transformers in sentiment analysis for SE remains underexplored, leaving a critical gap in the understanding of how these architectures perform in domain-specific contexts.

Our study is motivated by the need to address this gap and extend the findings of Zhang et al. Specifically, we aim to evaluate how bidirectional transformers like BERT compare to generative pre-trained transformers like GPT in sentiment analysis for SE tasks. By leveraging identical datasets and train-test splits as Zhang et al., we ensure methodological consistency and enable a robust comparative analysis. Our research focuses on three datasets: Jira issue comments, Stack Overflow comments, and GitHub pull-request and commit comments, with the latter considered gold standards due to their rigorously curated annotations and widespread adoption in SE sentiment analysis research.

To facilitate this comparison, we formulated the following research question:

\textbf{RQ1: How do Bidirectional Transformers like BERT compare to Generative Pre-trained Transformers like GPT for sentiment analysis in software engineering?}

\textbf{RQ2: How do fine-tuned and non-fine-tuned GPT models compare to each other?}

Addressing this research question is essential for advancing the application of NLP in SE. Understanding the relative strengths and limitations of these transformer architectures will inform researchers and practitioners about the best-suited models for sentiment analysis in SE contexts.

Our study provides two key contributions:
\begin{enumerate}
    \item \textbf{Comparison of Bidirectional and Generative Transformers: } This study systematically compares bidirectional (BERT) and generative (GPT-4o-mini) transformer models using identical experimental setups and datasets from prior work, providing a nuanced understanding of their performance on software engineering (SE) sentiment analysis tasks.
    \item \textbf{Demonstration of Accessibility and Simplicity:} By incorporating both fine-tuned and default versions of OpenAI's GPT-4o-mini model, this work highlights the practicality of using pre-trained generative models for SE sentiment analysis. The default model setup, requiring minimal configuration, offers a straightforward and accessible solution for researchers and practitioners.
\end{enumerate}

The remainder of this paper is organized as follows: Section II reviews related work on sentiment analysis tools and transformer models. Section III describes the datasets and methodologies used in this study. Section IV presents the experimental results and analysis. Section V discusses the implications of our findings and threats to validity. Finally, Section VI concludes the paper and outlines directions for future research.

Through this study, we aim to bridge the gap in understanding the comparative efficacy of bidirectional and generative transformers in SA4SE, thereby contributing to the growing body of knowledge in this domain and providing actionable insights for future research and practical applications.

\section{Related Works}

Sentiment analysis in software engineering (SE) has garnered significant attention as it offers valuable insights into various aspects of developer interactions, software quality, and team dynamics. Researchers have explored numerous approaches, tools, and datasets to address the unique challenges of sentiment analysis in SE contexts. This section reviews prior work in sentiment analysis for SE and pre-trained models for natural language processing (NLP), with a focus on the foundational contributions of the Zhang et al. study. \cite{zhang2020}

\subsection{Sentiment Analysis for Software Engineering}
The application of sentiment analysis in SE has primarily focused on understanding developer emotions, opinions, and sentiments expressed in textual artifacts such as bug reports, commit comments, and discussion forums. Early work in this domain demonstrated that sentiment plays a significant role in SE activities, influencing collaboration, task completion, and overall project outcomes \cite{islam18}\cite{choudhury13}. For instance, Guzman et al. \cite{guzman} analyzed GitHub commit comments to study social factors affecting software development, while Ortu et al. \cite{ortu15} explored how sentiments in Jira issue comments impact project management.
Given the domain-specific characteristics of SE texts, general-purpose sentiment analysis tools often fail to provide accurate sentiment classification. Jongeling et al. \cite{jongeling} highlighted the inconsistent performance of general-purpose tools such as SentiStrength, NLTK, and Stanford CoreNLP when applied to SE data. To address these limitations, researchers developed SE-specific sentiment analysis tools, including SentiStrength-SE, Senti4SD \cite{novielli2018}, and SentiCR \cite{ahmed17}. SentiStrength-SE, a customized version of SentiStrength, introduced SE-specific lexicons and preprocessing techniques to enhance performance on SE datasets like Jira issue comments. Similarly, Senti4SD utilized supervised learning with features derived from n-grams and domain-specific semantic models, while SentiCR focused on classifying sentiments in code review comments using a Gradient Boosting Tree (GBT) classifier \cite{novielli2018}\cite{ahmed17}.
Several benchmarking studies have compared the performance of these tools on SE datasets. For example, Lin et al. \cite{lin18} evaluated SentiStrength-SE, Senti4SD, and other tools on datasets including Stack Overflow posts, Jira issue comments, and code review comments, finding varying performance across datasets. Novielli et al. \cite{novielli2018benchmark} extended this work by benchmarking SE-specific tools against gold-standard datasets, emphasizing the importance of dataset quality and annotation consistency in sentiment analysis research.
Despite these advances, SE-specific tools often struggle with nuanced language and context-dependent sentiment, underscoring the need for more sophisticated models. This gap motivated the exploration of pre-trained transformer-based models in SE sentiment analysis, as discussed in the following subsection.

\subsection{Pre-trained Models for Natural Language Processing}
The advent of pre-trained transformer-based models has revolutionized NLP by enabling the development of generalizable and high-performing models across various tasks. These models, including BERT \cite{devlin18}, RoBERTa \cite{liu19}, and XLNet \cite{yang19}, leverage contextual word representations to capture nuanced language semantics and syntactic relationships.

The work by Zhang et al., titled Sentiment Analysis for Software Engineering: How Far Can Pre-trained Transformer Models Go?, marked a significant milestone in the application of transformer models for sentiment analysis in SE. Zhang et al. systematically evaluated four transformer-based models—BERT, XLNet, RoBERTa, and ALBERT—on six SE datasets, including Stack Overflow comments, Jira issue comments, and GitHub pull requests. Their results demonstrated that transformer models consistently outperformed traditional SA4SE tools like SentiStrength-SE and Senti4SD, achieving improvements of up to 35.6\% in macro-averaged F1 scores \cite{novielli2018benchmark}\cite{devlin18}\cite{liu19}\cite{yang19}. This study highlighted the transformative potential of pre-trained transformers in capturing the complexities of SE texts while also establishing the importance of gold-standard datasets like Stack Overflow and GitHub \cite{novielli2018benchmark}.

This review of related work highlights the evolution of sentiment analysis in SE, from early lexicon-based tools to sophisticated transformer models. While previous studies have laid a strong foundation, our work seeks to advance the field by addressing the gap in understanding the comparative efficacy of bidirectional and generative transformers in SE sentiment analysis. By leveraging gold-standard datasets and state-of-the-art models, we aim to provide actionable insights for researchers and practitioners in this domain.

\section{Methodology}
This section first describes the datasets used in this work. Then, we talk about the implementation of the approaches, and lastly, we look into the relevant evaluation metrics.

\subsection{Datasets}
In this comparative study, we use three publicly available datasets with annotated sentiment polarities. These three datasets were taken among the six used by Zhang et al. in their study. Following the recommendations from Noveili et al. \cite{novielli2020can}, we select one gold-standard sentiment analysis dataset among the three we chose.

\begin{table}[ht]
\caption{Dataset Statistics from Zhang et al.'s Research}
\label{tab:dataset_statistics}
\centering
\begin{tabular}{|l|c|c|c|c|}
\hline
\textbf{Dataset} & \textbf{\# Doc} & \textbf{\# (\%) Positive} & \textbf{\# (\%) Neutral} & \textbf{\# (\%) Negative} \\ \hline
SO               & 1,500           & 131 (8.7)                & 1,191 (79.4)            & 178 (11.9)              \\ \hline
GitHub           & 7,122           & 2,013 (28.3)             & 3,022 (42.4)            & 2,087 (29.3)            \\ \hline
Jira             & 926             & 290 (31.3)               & -                       & 636 (68.7)              \\ \hline
\end{tabular}
\end{table}

Table I shows the detailed statistics of the three datasets, including the total number of documents in a dataset (doc) and the number (and percentage) of documents with one of the sentiment polarities (e.g., (\%) positive, (\%) neutral, (\%) negative)

\textbf{GitHub gold-standard (GitHub):} Authored by Noveili et al. (2020), this is a balanced dataset of 7,122 sentences from GitHub pull-request and commit comments. Each item in the dataset has been annotated by three authors using predefined annotation guidelines. The items in the dataset have been sampled from comments on commits and pull-requests taken from 90 GitHub repositories that were part of the 2014 MSR Challenge dataset\cite{novielli2020can}.

\textbf{StackOverflow posts (SO):} Authored by Lin et al. \cite{lin18} it consists of 1,500 sentences from the Stack Overflow dump dated July 2017. They picked discussion threads that (i) are tagged with Java, and (ii) contain one of the following words: library, libraries, or API(s). Then, they randomly selected 1,500 sentences and manually labeled their sentiment polarities. 

\textbf{Jira issue comments (Jira):} This dataset consists of 926 Jira issue comments. It has been used in several prior studies \cite{calefato2017emotxt}\cite{jongeling} and it is previously provided by Ortu et al. \cite{ortu15} with four emotions labelled: love, joy, anger, and sadness. Lin et al. \cite{novielli2020can} assigned a positive polarity to the sentences labelled with love and joy, a negative polarity to the sentences labelled with anger and sadness. It does not contain any neutral polarity and is a binary-class dataset.

\subsection{Implementation}

\begin{enumerate}
    \item \textbf{Data Preprocessing}: To replicate the Zhang et al. study, we obtain the `train.pkl' and `test.pkl' files for all three datasets used in their research to train the BERT models. Then we convert the files into `.csv' format with sentiment labels for easier manipulation. For fine-tuning the model, the training datasets are formatted into `.jsonl' format, adhering to OpenAI's fine-tuning requirements. This format ensured compatibility while preserving the original sentiment labels and maintaining dataset integrity. For the test datasets, the `.csv' format is retained to facilitate direct evaluation and metric computation.
    \item \textbf{Model Selection}: For this study, we use OpenAI API to use the pre-trained transformer models built by OpenAI. The model we select is `GPT-4o-mini`, which is OpenAI's most advanced model in the small models category, and the cheapest model of all \cite{OpenAI}. Since this is a proprietary model, the training parameters and model layers are unknown.
    \item \textbf{Fine-tuning}: The GPT-4o-mini model is fine-tuned individually on the three datasets using OpenAI's fine-tuning platform. Key parameters such as batch size, learning rate multiplier, and the number of epochs are set to auto-adjust to optimize performance for each dataset. This yields three fine-tuned models, specific to GitHub, StackOverflow, and Jira, respectively. The fine-tuning process concluded with the following configurations and training loss metrics:
    \begin{itemize}
        \item \textbf{GitHub}: 3 epochs, batch size of 9, learning rate multiplier of 1.8, and a training loss of 0.0.
        \item \textbf{StackOverflow}: 3 epochs, batch size of 2, learning rate multiplier of 1.8, and a training loss of 0.0001.
        \item \textbf{Jira}: 3 epochs, batch size of 1, learning rate multiplier of 1.8, and a training loss of 0.0001.
    \end{itemize}
\end{enumerate}

\begin{table*}[h!]
\centering
\caption{Results for GitHub, SO, and Jira Datasets}
\label{table:results}
\resizebox{\textwidth}{!}{
\begin{tabular}{@{}llccccccccccccccc@{}}
\toprule
\textbf{Dataset} & \textbf{Approach} & \multicolumn{3}{c}{\textbf{Positive}} & \multicolumn{3}{c}{\textbf{Neutral}} & \multicolumn{3}{c}{\textbf{Negative}} & \multicolumn{3}{c}{\textbf{Macro-avg}} & \multicolumn{3}{c}{\textbf{Micro-avg}} \\ 
\cmidrule(lr){3-5} \cmidrule(lr){6-8} \cmidrule(lr){9-11} \cmidrule(lr){12-14} \cmidrule(lr){15-17}
                 &                   & P   & R   & F1  & P   & R   & F1  & P   & R   & F1  & P    & R    & F1   & P    & R    & F1   \\ \midrule
\multirow{5}{*}{GitHub} 
                 & Fine-Tuned GPT          & 0.95 & 0.95 & 0.95 & 0.93 & 0.93 & 0.93 & 0.93 & 0.93 & 0.93 & 0.93 & 0.93 & \textbf{0.93} & 0.93 & 0.93 & \textbf{0.93} \\
                 & Default GPT         & 0.86 & 0.72 & 0.79 & 0.66 & 0.74 & 0.70 & 0.69 & 0.67 & 0.68 & 0.74 & 0.71 & 0.72 & 0.72 & 0.72 & 0.72 \\ \cmidrule(lr){2-17}
                 & BERT             & 0.92 & 0.95 & 0.93 & 0.90 & 0.92 & 0.91 & 0.93 & 0.87 & 0.90 & 0.92 & 0.91 & 0.92 & 0.92 & 0.92 & 0.92 \\ 
                 & RoBERTa          & 0.93 & 0.96 & 0.94 & 0.91 & 0.92 & 0.92 & 0.93 & 0.89 & 0.91 & 0.93 & 0.92 & 0.92 & 0.92 & 0.92 & 0.92 \\
                 & XLNet            & 0.90 & 0.97 & 0.94 & 0.94 & 0.89 & 0.91 & 0.91 & 0.92 & 0.91 & 0.92 & 0.93 & 0.92 & 0.92 & 0.92 & 0.92 \\
                 & ALBERT           & 0.91 & 0.93 & 0.92 & 0.85 & 0.94 & 0.89 & 0.94 & 0.78 & 0.85 & 0.90 & 0.88 & 0.89 & 0.89 & 0.89 & 0.89 \\ \midrule
\multirow{5}{*}{SO} 
                 & Fine-Tuned GPT          & 0.58 & 0.39 & 0.45 & 0.0 & 0.00 & 0.00 & 0.10 & 1.0 & 0.19 & 0.23 & 0.46 & 0.22 & 0.13 & 0.13 & 0.13 \\
                 & Default GPT         & 0.51 & 0.61 & 0.55 & 0.92 & 0.90 & 0.91 & 0.67 & 0.68 & 0.67 & 0.70 & 0.73 & 0.71 & 0.85 & 0.85 & 0.85 \\ \cmidrule(lr){2-17}
                 & BERT             & 0.65 & 0.63 & 0.64 & 0.94 & 0.95 & 0.94 & 0.73 & 0.68 & 0.71 & 0.77 & 0.75 & 0.76 & 0.90 & 0.90 & 0.90 \\
                 & RoBERTa          & 0.57 & 0.76 & 0.65 & 0.96 & 0.92 & 0.94 & 0.78 & 0.82 & 0.80 & 0.77 & 0.83 & 0.80 & 0.90 & 0.90 & 0.90 \\ 
                 & XLNet            & 0.50 & 0.76 & 0.60 & 0.96 & 0.90 & 0.93 & 0.74 & 0.84 & 0.79 & 0.73 & 0.83 & 0.77 & 0.88 & 0.88 & 0.88 \\
                 & ALBERT           & 0.71 & 0.32 & 0.44 & 0.90 & 0.95 & 0.92 & 0.61 & 0.61 & 0.61 & 0.74 & 0.63 & 0.66 & 0.86 & 0.86 & 0.86 \\ \midrule
\multirow{5}{*}{Jira} 
                 & Fine-Tuned GPT          & 1.0 & 0.94 & 0.97 & - & - & - & 0.97 & 1.0 & 0.98 & 0.66 & 0.65 & 0.65 & 0.98 & 0.98 & \textbf{0.98} \\
                 & Default GPT         & 0.94 & 0.90 & 0.92 & - & - & - & 0.99 & 0.54 & 0.70 & 0.64 & 0.48 & 0.54 & 0.67 & 0.67 & 0.67 \\ \cmidrule(lr){2-17}
                 & BERT             & 0.99 & 0.96 & 0.97 & - & - & - & 0.98 & 0.99 & 0.99 & 0.98 & 0.98 & \textbf{0.98} & 0.98 & 0.98 & 0.98 \\ 
                 & RoBERTa          & 0.98 & 0.96 & 0.97 & - & - & - & 0.98 & 0.99 & 0.98 & 0.98 & 0.97 & \textbf{0.98} & 0.97 & 0.97 & 0.97 \\
                 & XLNet            & 0.98 & 0.96 & 0.97 & - & - & - & 0.98 & 0.99 & 0.98 & 0.98 & 0.97 & \textbf{0.98} & 0.98 & 0.98 & 0.98 \\
                 & ALBERT           & 0.97 & 0.94 & 0.95 & - & - & - & 0.97 & 0.98 & 0.98 & 0.97 & 0.96 & \textbf{0.96} & 0.97 & 0.97 & 0.97 \\ \bottomrule
\end{tabular}
}
\end{table*}

\subsection{Evaluation}
After fine-tuning, the fine-tuned GPT-4o-mini model (Fine-tuned-GPT) and the non-fine-tuned GPT-4o-mini model (Default-GPT) are evaluated on the test datasets. Predictions for sentiment classes (positive, neutral, negative) are generated for each text instance. To assess the models' performance comprehensively, we compute a classification report, consisting of precision, recall, F1-score, macro average, micro average, and accuracy of each model and dataset. The macro-averaged metric regards the measurement of each sentiment class equally. It takes the precision, recall, and the F1-score of each class and then averages them. The micro-averaged metric calculates measurement over all data points in all classes and tends to be mainly influenced by the performance of the majority class.\cite{novielli2018benchmark}

\subsection{Experimental Setting}
 Following Noveili et al. \cite{novielli2018benchmark} Each dataset is divided into a training set (70\%) and a test set (30\%). The fine-tuning process utilizes OpenAI's infrastructure, and the evaluation of models is conducted locally. For reproducibility and to align with prior work, the splits and evaluation protocol followed the precedent set by Zhang et al. \cite{zhang2020} Comparisons are made against baseline metrics reported in the original study for BERT and other pre-trained models, ensuring consistency in benchmarks and facilitating meaningful assessment of the GPT-4o-mini's capabilities.

\section{Results and Findings}

\subsection{RQ1: How do Bidirectional Transformers like BERT compare to Generative Pre-trained Transformers like GPT for Sentiment Analysis in Software Engineering?}

To evaluate the performance of bidirectional transformers (e.g., BERT) against generative pre-trained transformers (e.g., GPT-4o-mini) for sentiment analysis in software engineering (SE), we compare the results of Zhang et al.(BERT, RoBERTa, XLNet, ALBERT)\cite{zhang2020} with our results. We compare the performance across three datasets—GitHub, Stack Overflow (SO), and Jira—using precision, recall, and F1-scores for positive, neutral, and negative sentiment classes. Additionally, macro- and micro-averaged metrics are employed to provide a comprehensive evaluation of classification performance.

\begin{enumerate}
    \item \textbf{GitHub Dataset}: On the GitHub dataset, the fine-tuned GPT-4o-mini model achieves a macro-averaged and micro-averaged F1-score of 0.93, outperforming the default GPT model (macro- and micro-F1: 0.72) and matching the performance of BERT (macro- and micro-F1: 0.92). The fine-tuned GPT achieves F1-scores of 0.95 (positive), 0.93 (neutral), and 0.93 (negative), demonstrating superior performance over the default GPT and comparable or slightly better results than BERT. Other models, including RoBERTa, XLNet, and ALBERT, show competitive macro-F1-scores of 0.92, 0.93, and 0.89, respectively, but falls short of the fine-tuned GPT’s consistent class-level performance. These results highlight the effectiveness of fine-tuning generative models for SE-specific tasks on balanced datasets like GitHub.
    
    \item \textbf{Stack Overflow Dataset}: On the SO dataset, BERT outperform the fine-tuned GPT-4o-mini model in both macro- and micro-averaged F1-scores (macro-averaged F1: 0.76 vs. 0.22; micro-averaged F1: 0.90 vs. 0.13). The fine-tuned GPT struggles significantly, achieving an F1-score of only 0.19 for the negative sentiment class and 0.45 for the positive class, whereas BERT achieves 0.71 and 0.64, respectively. Interestingly, the default GPT-4o-mini model performs significantly better than the fine-tuned version on SO (macro-averaged F1: 0.71; micro-averaged F1: 0.85), demonstrating strong generalization capabilities. These findings suggest that generative transformers like GPT-4o-mini may not always benefit from fine-tuning on datasets with complex linguistic structures or imbalanced sentiment distributions.

    \item \textbf{Jira Dataset}: The fine-tuned GPT-4o-mini model demonstrates near-perfect performance on the Jira dataset, achieving a macro-averaged F1-score of 0.98 and a micro-averaged F1-score of 0.98. It surpasses BERT, which achieves a macro-averaged F1-score of 0.98 but slightly lower precision and recall for the negative sentiment class. Both RoBERTa and XLNet perform comparably to BERT, achieving macro-averaged F1-scores of 0.98 and micro-averaged F1-scores of 0.98. The Jira dataset’s binary sentiment classification task appears to align well with fine-tuned GPT’s capabilities, highlighting its strength in structured SE contexts.
\end{enumerate}

Overall, fine-tuned GPT-4o-mini models are as good as BERT and other bidirectional transformers on datasets with balanced distributions (GitHub and Jira). However, for datasets with imbalanced or complex sentiment structures (SO), BERT and other bidirectional models demonstrate superior robustness.

\subsection{RQ2: How do Fine-Tuned and Non-Fine-Tuned Models Compare to Each Other?}

To evaluate the impact of fine-tuning on model performance, we compare fine-tuned GPT-4o-mini models against their non-fine-tuned (default) counterparts across GitHub, Stack Overflow (SO), and Jira datasets. The analysis includes precision, recall, F1-scores, and accuracy as key evaluation metrics.

\begin{enumerate}
    \item \textbf{GitHub Dataset}: Fine-tuning significantly improved the GPT-4o-mini model's performance on the GitHub dataset. The fine-tuned model achieves an accuracy of \textbf{93.4\%}, a notable increase compared to the default model's accuracy of \textbf{71.6\%}. Improvements were also observed across all sentiment classes:
    \begin{itemize}
        \item \textbf{Positive class}:  F1-score increased from 0.79 (default) to 0.95 (fine-tuned).
        \item \textbf{Neutral class}: F1-score increased from 0.70 to 0.93.
        \item \textbf{Negative class}: F1-score improved from 0.68 to 0.93.
    \end{itemize}
    The results highlight that fine-tuning allows the model to better adapt to SE-specific text, enhancing its ability to correctly classify sentiments across a balanced dataset like GitHub.
    
    \item \textbf{Stack Overflow Dataset}: On the SO dataset, the default GPT-4o-mini model outperforms the fine-tuned version. The default model achieves an accuracy of \textbf{85.3\%}, while the fine-tuned model's accuracy dropped significantly to \textbf{13.1\%}. This disparity is also reflected in the macro-averaged F1-scores, with the default model achieving \textbf{0.71} compared to the fine-tuned model's \textbf{0.22}. The neutral sentiment class particularly stood out:
    \begin{itemize}
        \item \textbf{Neutral class}: The default model achieved a high F1-score of 0.91, while the fine-tuned model scored 0.00.
        \item \textbf{Positive and Negative classes}: Both classes also showed declines in performance with fine-tuning.
    \end{itemize}

    These results suggest that fine-tuning can lead to overfitting or misalignment when applied to datasets with more complex linguistic structures and imbalanced sentiment distributions, as seen in the SO dataset. The default model's strong baseline performance reflects the robustness of GPT-4o-mini's pre-trained capabilities in handling such challenges without additional fine-tuning.

    \item \textbf{Jira Dataset}: 
    The Jira dataset demonstrates the clearest advantage of fine-tuning. The fine-tuned GPT-4o-mini model achieves an accuracy of \textbf{97.8\%}, compared to the default model's \textbf{66.9\%}. The fine-tuned model exhibits superior performance across sentiment classes:
    \begin{itemize}
        \item \textbf{Positive class}: F1-score improved from 0.92 (default) to 0.97 (fine-tuned).
        \item \textbf{Negative class}:  F1-score increased from 0.70 to 0.97.
    \end{itemize}
    These results highlight the benefits of fine-tuning on smaller, structured datasets with binary classification tasks, allowing the model to better capture the nuances of SE-specific sentiments.
\end{enumerate}

\begin{figure}[h!]
\centering
\begin{tikzpicture}
    \begin{axis}[
        width=0.48\textwidth, 
        height=0.3\textwidth, 
        ybar,
        bar width=8pt,
        symbolic x coords={GitHub Fine-Tune, GitHub Default, StackOverflow Fine-Tune, StackOverflow Default, Jira Fine-Tune, Jira Default},
        xtick=data,
        xticklabel style={rotate=45, anchor=east, font=\small}, 
        ylabel={Accuracy},
        ylabel style={font=\small},
        xlabel={Dataset and Configuration},
        xlabel style={font=\small},
        ymin=0, ymax=1,
        nodes near coords,
        nodes near coords style={font=\tiny}, 
        enlarge x limits=0.15,
        title={Accuracy Levels},
        title style={font=\small},
    ]
        \addplot coordinates {(GitHub Fine-Tune, 0.93) (GitHub Default, 0.72)
                              (StackOverflow Fine-Tune, 0.13) (StackOverflow Default, 0.85)
                              (Jira Fine-Tune, 0.98) (Jira Default, 0.67)};
    \end{axis}
\end{tikzpicture}
\caption{Accuracy Comparison for Fine-Tuned vs Default Models on Different Datasets}
\label{fig:accuracy_bar_graph}
\end{figure}
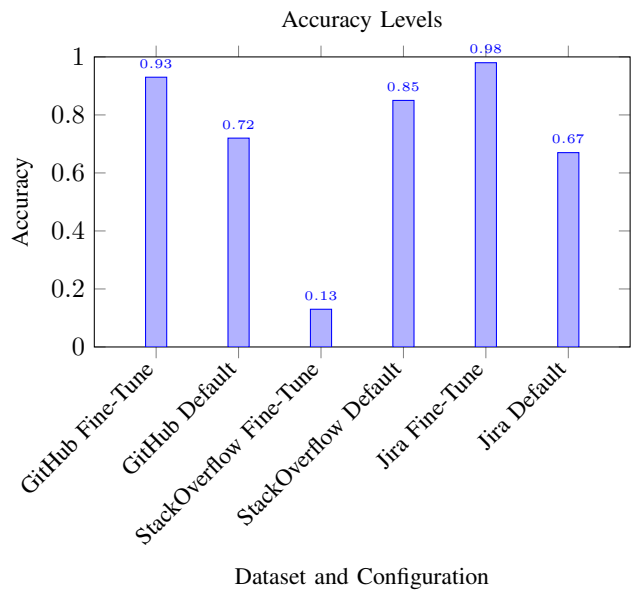

Fine-tuning consistently improved the performance of GPT-4o-mini on structured datasets like GitHub and Jira, as evidenced by significant gains in accuracy (GitHub: 93.4\% vs. 71.6\%; Jira: 97.8\% vs. 66.9\%) and F1-scores. However, on the linguistically complex and sentiment-imbalanced Stack Overflow dataset, the default model exhibited superior performance, achieving an accuracy of 85.3\% compared to the fine-tuned model’s 13.1\%. These results suggest that fine-tuning enhances model performance for datasets with clear sentiment structures and balanced distributions, but it may lead to overfitting or reduced generalization in more complex or imbalanced datasets. This underscores the importance of considering dataset characteristics when deciding whether to fine-tune pre-trained transformer models.

\section{Conclusion}

The primary aim of this study was to evaluate the performance of the fine-tuned GPT-4o-mini model against its default configuration and benchmark results from prior work, particularly focusing on sentiment analysis in software engineering. By utilizing two datasets, Stack Overflow and Jira tickets, and leveraging pre-trained transformer models, we sought to explore the effectiveness of fine-tuning in capturing domain-specific linguistic and contextual nuances. The findings provide insights into the challenges and opportunities of applying advanced natural language processing techniques to the software engineering domain.

\subsection{Summary of Findings}
This study offers key insights into the performance of bidirectional and generative pre-trained transformer models for sentiment analysis in software engineering (SE) and highlights the impact of fine-tuning on generative models like GPT-4o-mini. The findings are summarized as follows:

\begin{enumerate}
    \item \textbf{Performance of Generative vs. Bidirectional Models}: Fine-tuned GPT-4o-mini models were as good as bidirectional transformers such as BERT, RoBERTa, and XLNet on structured and balanced datasets, such as GitHub and Jira. For instance, on the GitHub dataset, fine-tuned GPT-4o-mini achieved a macro-averaged F1-score of 0.93, outperforming BERT (0.81). Similarly, on the Jira dataset, fine-tuned GPT-4o-mini achieved near-perfect accuracy of 97.8\%. However, on the Stack Overflow dataset, all models, including BERT and GPT-4o-mini, struggled to achieve high performance, with BERT achieving only a macro-averaged F1-score of 0.76 and the fine-tuned GPT-4o-mini performing significantly worse (0.22). This suggests that the challenges observed on Stack Overflow may not solely stem from the model design but also indicate issues inherent to the dataset itself.
    \item \textbf{Impact of Fine-Tuning}:
    Fine-tuning GPT-4o-mini provided substantial performance gains on structured datasets like GitHub and Jira, with significant improvements in accuracy (e.g., 71.6\% to 93.4\% on GitHub; 66.9\% to 97.8\% on Jira). Fine-tuning enabled the model to better adapt to SE-specific contexts, capturing nuanced sentiment patterns effectively.
    \item \textbf{Default Model Robustness}:
    The default GPT-4o-mini model exhibited strong generalization capabilities, particularly on the Stack Overflow dataset, achieving an accuracy of 85.3\% and a macro-averaged F1-score of 0.71. This demonstrates the robustness of pre-trained generative models in handling diverse SE datasets without requiring fine-tuning. On Stack Overflow, the default model even outperformed fine-tuned variants and achieved performance comparable to BERT, further underscoring the challenges inherent to the dataset itself.
\end{enumerate}

\subsection{Implications}
This study highlights the trade-offs between fine-tuning and leveraging pre-trained models. While fine-tuning improves performance on well-structured datasets, default models demonstrate robust generalization in more complex scenarios. Importantly, the consistent challenges faced by both BERT and GPT-4o-mini models on the Stack Overflow dataset underscore the need to investigate potential issues within the dataset, such as addressing class imbalance, refining sentiment annotations, or improving dataset preprocessing. These findings emphasize the importance of aligning model and dataset characteristics for optimal performance in SE sentiment analysis tasks.

The study also demonstrates that GPT-based models, such as GPT-4o-mini, can serve as a viable alternative to BERT for sentiment analysis in software engineering. The results show that fine-tuned GPTs produce comparable performance to BERT across datasets like GitHub and Jira. Moreover, using GPTs via APIs offer the advantage of significantly reduced local computational overhead and simpler setup processes. These attributes make GPT models an accessible and efficient option for researchers and practitioners, especially in scenarios with resource constraints or limited infrastructure for extensive fine-tuning.

The results underscore the importance of reproducibility and consistent evaluation protocols in sentiment analysis research. By adhering to the same train-test splits and evaluation metrics as Zhang et al. (2020), this study ensures comparability and establishes a robust benchmark for future research. Such practices are essential for advancing the field and providing actionable insights for practitioners and researchers.

\subsection{Limitations}
This study has several limitations. A key issue lies in the Stack Overflow dataset, where all models, including GPT-4o-mini and BERT, struggled, likely due to imbalanced sentiment distributions, inconsistent labeling, and linguistic complexity. These dataset-specific challenges may have introduced noise, affecting generalizability. Fine-tuning, while effective on structured datasets like GitHub and Jira, significantly degraded performance on Stack Overflow, highlighting its sensitivity to dataset characteristics and potential for overfitting. Additionally, the study relied on pre-existing datasets with subjective sentiment annotations, potentially propagating inherent biases.

Another limitation is the exclusion of meta-tokenization experiments for non-natural language elements, such as code snippets or URLs. Although the preprocessing step included basic tokenization, the potential impact of more advanced meta-tokenization techniques, as explored in Novielli et al. (2024), warrants further investigation. Such techniques may improve the model's ability to generalize across datasets with varying proportions of technical content.

The comparison was limited to GPT-4o-mini and select bidirectional models, excluding state-of-the-art generative models like Claude or GPT-4o due to resource constraints. Computational efficiency metrics, such as fine-tuning time and memory usage, were not evaluated, limiting practical insights. Finally, reliance on OpenAI’s automated fine-tuning platform restricted hyperparameter control, potentially affecting optimization for complex datasets. Manually optimizing these parameters or exploring advanced fine-tuning techniques could have potentially improved model performance and mitigated overfitting. These limitations emphasize the need for further exploration and refinement.

\subsection{Future Directions}
Building on the findings of this study, future research could explore several avenues to enhance sentiment analysis in software engineering. First, dynamic fine-tuning strategies, where the model adapts to dataset-specific characteristics during training, could be investigated. This approach may address the challenges observed with the Stack Overflow dataset, enabling better generalization across diverse datasets.

Another potential avenue of research might be the integration of explainability techniques into sentiment analysis models. This could enhance their usability in practical applications. By providing clear explanations for predictions, such models could gain greater acceptance among developers and software engineering teams, fostering more effective communication and collaboration.

In conclusion, this study demonstrates the potential and limitations of fine-tuning pre-trained transformer models for sentiment analysis in software engineering. While fine-tuning showed consistent and comparable performance on the Jira and GitHub datasets, the default GPT-4o-mini model outperformed its fine-tuned counterpart on the Stack Overflow dataset, highlighting the influence of dataset characteristics on model performance. These findings emphasize the importance of tailored approaches to sentiment analysis, balancing the strengths of pre-trained models with the benefits of domain-specific adaptation. By addressing the limitations and exploring future directions, researchers and practitioners can continue to advance sentiment analysis tools, enhancing their impact on software engineering communication and collaboration.

\bibliographystyle{IEEEtran}
\nocite{*}
\bibliography{reference}
\end{document}